\documentclass{ws-procs975x65}

\begin{document}

\title{A REVIEW OF THE LARGE $N$ LIMIT OF TENSOR MODELS}

\author{R. GURAU $^*$}

\address{Perimeter Institute, 31 Caroline St. N., Waterloo, ON, N2L 2Y5, Canada\\
$^*$E-mail: rgurau@perimeterinstitute.ca }

\begin{abstract}

Random matrix models encode a theory of random two dimensional surfaces
with applications to string theory, conformal field theory, statistical physics in
random geometry and quantum gravity in two dimensions. The key to their success lies 
in the $1/N$ expansion introduced by 't Hooft. Random tensor models generalize
random matrices to theories of random higher dimensional spaces.
For a long time, no viable $1/N$ expansion for tensors was known and their 
success was limited.

A series of recent results has changed this situation and the extension of the
$1/N$ expansion to tensors has been achieved. We review these results in this paper.
\end{abstract}


\bodymatter

\section{Introduction}\label{aba:sec1}

In many theories space appears as a fixed background, a manifold with a fixed metric. Length scales are defined
with respect to this background. The scales encode the causality: fundamental physics at short distances 
determines the effective phenomena at large distances.
It is however clear that the length scale are too crude to be applied in all instances: not only general relativity promotes the metric to 
a dynamical variable but also quantum field theory suggests that at the fundamental
level space should become random,  quantum. A fundamental theory of space is for now out of reach. Nevertheless we know that, no matter what this
theory is, it must answer three major questions:
how to define a statistical theory of random geometries, how to define an appropriate notion of scale and how to recover
the usual space time as an effective phenomenon. 

In this paper we present a framework to deal with random geometries in arbitrary dimensions.
In this framework we answer precisely each of the three question formulated above. 

Invariant probability distributions for random $N \times N$ matrices\footnote{Random matrices were introduced by Wishart \cite{wishart} in statistics and 
used for the first time in physics by Wigner \cite{wigner} for the study of the spectra of heavy atoms.} 
encode the most successful (but restricted to two dimensions) theory of random geometry we have so far. The moments and 
partition functions of such a probability distribution evaluate in terms of ribbon Feynman graphs.
Ribbon graphs are in one to one correspondence with triangulated surfaces\cite{Di Francesco:1993nw}
providing the bridge between random matrices and random geometries.

In his seminal work \cite{'tHooft:1973jz} 't Hooft realized that matrix models have a built in notion of scale: the size of the matrix, $N$.
The perturbative expansion can be reorganized in powers of $1/N$ (indexed by the genus) and in the large $N$ 
limit only planar graphs \cite{Brezin:1977sv} contribute. Matrix models
undergo a phase transition to continuum infinitely refined surfaces\cite{Kazakov:1985ds,mm} because 
the planar graphs form a summable family (i.e. a power series with a finite radius of convergence). 
Random matrices provide the quantization of two dimensional 
gravity coupled to conformal matter \cite{Kazakov:1986hu, Boulatov:1986sb, Brezin:1989db,Kazakovmulticrit,
Ambjorn:1990ji,Fukuma:1990jw,Makeenko:1991ry,Dijkgraaf:1990rs} and through the KPZ correspondence 
\cite{Knizhnik:1988ak, david2, DK, Dup} they relate to conformal field theory in fixed geometry.
In the double scaling limit matrix models provide a quantization of the string 
world sheet\cite{double,double1,double2} in string theory.

The resounding success of matrix models has inspired their generalization in higher dimensions 
to random tensor models\cite{ambj3dqg,sasa1,mmgravity,sasab,sasac,Oriti:2011jm,Baratin:2011aa}.
Until recently however tensor models have failed 
to provide an analytically controlled theory as, for a long time,
no generalization of the $1/N$ expansion to tensors has been found. 

The discovery of colored\cite{color,PolyColor,lost} tensor 
models\cite{coloredreview,sefu2,Ryan:2011qm,Carrozza:2011jn,Carrozza:2012kt,IsingD,EDT,doubletens,Bonzom:2012sz,Bonzom:2012qx} 
has drastically changed this situation. The colored models support a $1/N$ expansion\cite{Gur3,GurRiv,Gur4} indexed by the {\it degree}. 
The leading order (melonic\cite{Bonzom:2011zz}) graphs, triangulate the sphere in any dimension\cite{Gur3,GurRiv,Gur4}. 
They map to trees\cite{Bonzom:2011zz} hence are a summable family (different form the planar family).
Colored tensor models undergo a phase transition\cite{Bonzom:2011zz} to a theory of continuous infinitely refined random spaces. 

These results generalize to all invariant models for a random complex tensor\cite{Bonzom:2012hw}. 
The colors arise as a canonical bookkeeping device labeling the indices of the tensor.
The resulting universal\cite{Gurau:2011kk,Gurau:2011tj,Gurau:2012ix,Bonzom:2012fu} theory of random tensors 
is the generalization of invariant matrix models to higher dimensions. The $1/N$ expansion can be realized 
dynamically as a renormalization group flow in tensor group field theories\cite{BenGeloun:2011rc,BenGeloun:2012pu,
BenGeloun:2012yk,Carrozza:2012uv,Geloun:2012bz,Rivasseau:2011hm} which are, in at least two simple cases, asymptotically free\cite{BenGeloun:2012pu,BenGeloun:2012yk}.

\section{Tensor Models}

Let us consider a covariant complex tensor of rank $D$ transforming under the external tensor product of $D$ fundamental 
representations of the unitary group $\otimes_{i=1}^D U(N_i)$
\begin{eqnarray}
&& T'_{a_1\dotsc a_D} = \sum_{n_1,\dotsc,n_D} U_{a_1n_1}\dotsm V_{a_Dn_D}\ T_{n_1\dotsc n_D}  \crcr
&& \bar T'_{  a_1\dots  a_D} = \sum_{n_1,\dotsc,n_D} \bar U_{  a_D n_D}\dotsm \bar V_{a_1 n_1}\ \bar T_{n_1\dots n_D}  \; .
\end{eqnarray} 
We stress that each unitary group $U(N_i)$ acts independently on its corresponding index. The tensor 
$T$ has {\it no symmetry properties} under permutation of its indices. The dimensions $N_1, \dots N_D$ can be different
but for simplicity we set $N_i =N$, $\forall i$.
We denote by $\bar n_i$ the indices of the complex conjugated tensor $\bar T$  (which is contravariant of rank $D$),
and by $\vec n$ the $D$-uple of integers $(n_1, \dotsc, n_D)$.
We define the {\it color} of an index as its position: $n_1$ (and $\bar n_1$) has color $1$, $n_2$ (and $\bar n_2$) has color $2$ and so on. 

\paragraph{Invariants and action.}

An invariant tensor model is a probability measure defined by an invariant polynomial
$S (T,\bar  T) $, which we call the ``action'',
\begin{equation} 
 d\nu= \frac{1}{Z } e^{  - N^{D-1} S (T,\bar  T)  } \; \Bigl( \prod dT_{n_1 \dots n_D} 
 d\bar T_{ n_1 \dots n_D } \Bigr)\;.
\end{equation}

By the fundamental theorem of classical invariants of the unitary group
(see\cite{collins} for a modern proof) any invariant polynomial
is a linear combination of {\it trace invariants} obtained by contracting 
pairs of covariant and contravariant indices in a product of tensor entries
\begin{eqnarray}
\text{Tr} (T,\bar T) = \sum \prod \delta_{n_1,\bar n_1} \;  T_{n_1\dotsc} \dots \bar T_{\bar n_1 \dots } \; ,
\end{eqnarray} 
where all indices are saturated. Note that, in order to obtain an invariant,
we always contract the first index $n_1$ of a $T$ with the first index $\bar n_1$ on some $\bar T$, the second index $n_2$ of 
$T$ with the second index $\bar n_2$ on some $\bar T$ and so on. That is we only contract indices of the {\it same color}.

\begin{figure}[t]
\begin{center}
 \includegraphics[width=5cm]{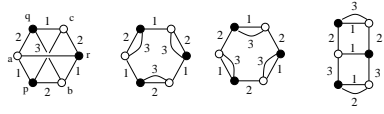}  
\caption{Some trace invariants for $D=3$.}
\label{fig:tensobs}
\end{center}
\end{figure}

The trace invariants can be represented graphically as $D$-colored graphs.
We represent every $T_{\dots n_i \dots}$ by a white vertex $v$ and every $\bar T_{\dots n_i \dots}$ by a black vertex $\bar v$.
The contraction of the two indices $n_i$ and $\bar n_i$, $\delta_{ n_i \bar n_i}$, is represented by a 
line $l^i = (v,\bar v)$ of color $i$ connecting the two vertices (see figure \ref{fig:tensobs}). For example the 
invariant 
\begin{eqnarray}
        && \sum \delta_{a_1p_1} \delta_{a_2q_2} \delta_{a_3r_3} \quad   
                    \delta_{b_1r_1} \delta_{b_2p_2} \delta_{b_3q_3} \quad  
                   \delta_{c_1q_1} \delta_{c_2r_2} \delta_{c_3p_3}  \crcr
                && \qquad T_{a_1a_2a_3}  T_{b_1b_2b_3} T_{c_1c_2c_3}  
           \bar T_{p_1 p_2p_3} \bar T_{q_1 q_2q_3} \bar T_{r_1 r_2r_3} \; ,
\end{eqnarray}
is represented by the leftmost graph in figure \ref{fig:tensobs} (the vertex $a$ in the drawing represents $T_{a_1a_2a_3}$ 
and so on). The trace invariant associated to the graph ${\cal B}$ writes
\begin{eqnarray}
 \text{Tr}_{{\cal B}}(T,\bar T ) = \sum_{\{\vec n^v,\bar{\vec{n}}^v\}_{v,\bar v \in {\cal B}}} 
  \Bigl( \prod_{i=1}^D \prod_{l^i = (v,\bar v)\in {\cal B}} \delta_{n_i^v \bar n_i^{\bar v}} \Bigr)  \; 
\prod_{v,\bar v \in {\cal B}} T_{\vec n^v} \bar T_{\bar {\vec n }^{\bar v} } \;,
\end{eqnarray}
where $v$ and $\bar v$ run over the white and black vertices of ${\cal B}$ and $l^i$ runs over the lines of color $i$ of ${\cal B}$. 
The (unique) graph with two vertices connected by $D$ lines is called the $D$-dipole and is denoted ${\cal B}_1$. It represents the 
unique quadratic invariant
\begin{eqnarray}\label{eq:gaussian}
 \text{Tr}_{{\cal B}_1} (T , \bar T ) = \sum_{\vec n,\bar{\vec{n}}}\, T_{\vec n}\, \bar T_{\bar {\vec n} }\ \Bigl[\prod_{i=1}^D \delta_{n_i \bar n_i}\Bigr] \; .
\end{eqnarray} 
The most general ``single trace'' invariant tensor model is defined by the action 
\begin{eqnarray} \label{eq:actiongen}
 S(T,\bar T) =  \, \text{Tr}_{{\cal B}_1} (T , \bar T ) + \sum_{{\cal B} } t_{{\cal B}}\, N^{-\frac{2}{(D-2)!} \omega({\cal B})} \,
\text{Tr}_{{\cal B}}(T,\bar T)\;,
\end{eqnarray}
where the sum over invariants includes only {\it connected} graphs ${\cal B}$.
The parameters $t_{\cal B}$ are the coupling constants of the model.
We have added in eq. \eqref{eq:actiongen} a scaling $ N^{-\frac{2}{(D-2)!} \omega({\cal B})} $ for each invariant 
as it simplifies some formulae. Up to trivial modifications all the results presented below hold in its absence.

\paragraph{Feynman graphs.}

We will discuss in the sequel the partition function 
\begin{equation} 
Z(t_{\cal B}) = \int \Bigl( \prod_{\vec n =  \vec {\bar n} } dT_{\vec n} d\bar T_{\vec { \bar n} } \Bigr)\;   e^{ -N^{D-1} S (T,\bar  T)  } \;,
\end{equation}
as the expectations of the invariant observables (which are also trace invariants represented by $D$-colored graphs)
are treated similarly.

\begin{figure}[t]
\begin{center}
 \includegraphics[width=2.5cm]{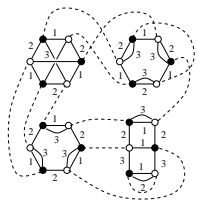}  
\caption{A Feynman graph.}
\label{fig:tensobsgraph}
\end{center}
\end{figure}

By Taylor expanding with respect to $t_{{\cal B}}$ and evaluating the Gaussian integral in terms of Wick contractions,
$Z(t_{\cal B}) $ becomes a sum over Feynman graphs. The graphs are made of {\it effective vertices} connected by
\emph{effective  propagators}. The effective vertices are the invariants $ \text{Tr}_{{\cal B}}(T,\bar T) $, 
that is they are themselves graphs ${\cal B}$ with colors $1,\dotsc, D$.
The effective propagators correspond to pairings (Wick contractions) of $T_{a_1\dotsc a_D}$'s 
and $\bar T_{\bar p_1 \dotsc \bar p_D}$'s. 
A Wick contraction with the quadratic part eq. \eqref{eq:gaussian} amounts to replacing the two tensors by 
$ \frac{ 1}{N^{D-1} } \prod_{i=1}^D \delta_{a_i \bar p_i} $. We represent the effective propagators
by dashed lines to which we assign a new color, $0$. 
The Feynman graphs, denoted from now on $ {\cal G} $, are then $D+1$ colored graphs 
(see figure \ref{fig:tensobsgraph}). 

We label ${\cal B}_{(\rho)}, \; \rho = 1, \dots |\rho| $ the effective vertices (subgraphs with colors $1,\dots D$) 
of ${\cal G} $ and 
denote $l^0$ the effective propagators (lines of color $0$) of ${\cal G}$.
The free energy is a sum over connected Feynman graphs ${\cal G}$
\begin{equation}
F(t_{{\cal B}}) = -\ln Z( t_{{\cal B}})  = \sum_{{\cal G} } \frac{ (-1)^{ |\rho| }}{s({\cal G})}
\ \Bigl( \prod_{\rho=1}^{|\rho|} t_{{\cal B}_{(\rho)}} \Bigr) \, A({\cal G}) \;,
\end{equation}
where $s({\cal G})$ is a symmetry factor and $A({\cal G})  $ is the amplitude of ${\cal G}$
\begin{eqnarray}\label{eq:ampli}
A({\cal G}) &=&  \sum_{\{\vec{n}^v,\bar{\vec{n}}^{\bar v}\}}\, \Bigl[
\prod_{\rho}  N^{D-1 -\frac{2}{(D-2)!} \omega({\cal B}_{(\rho)} )   } 
\Bigl( \prod_{i=1}^D \prod_{l^i = (v,\bar v)\in {\cal B_{(\rho)}}} \delta_{n_i^v \bar n_i^{\bar v}} \Bigr)  \Bigr]
\crcr 
 && \qquad \Bigl[\prod_{l^0=(v,\bar v) } \frac{1}{  N^{D-1}} \prod_{i} \delta_{n_i^v,\bar n_{i}^{\bar v}} \Bigr]\; .
\end{eqnarray}

\paragraph{Colored Graphs and triangulated spaces.}

Matrix models generate ribbon graphs which represent triangulated surfaces. 
The colored graphs of tensor models represent triangulated spaces\cite{lost}:

\begin{theorem}\label{th:pseudo}
  A closed connected $D+1$ colored graph is a $D$ dimensional normal simplicial pseudo manifold.
\end{theorem}

A pseudo manifold is a generalization of a manifolds having a finite number of conical singularities. 
This pseudo manifold can be built by gluing simplices. Consider $D=3$, thus  
the $D+1$ colored graphs have lines of colors $0$, $1$, $2$ and $3$. 

\def\figsubcap#1{\par\noindent\centering\footnotesize(#1)}
\begin{figure}[ht]%
\begin{center}
  \parbox{1cm}{\epsfig{figure=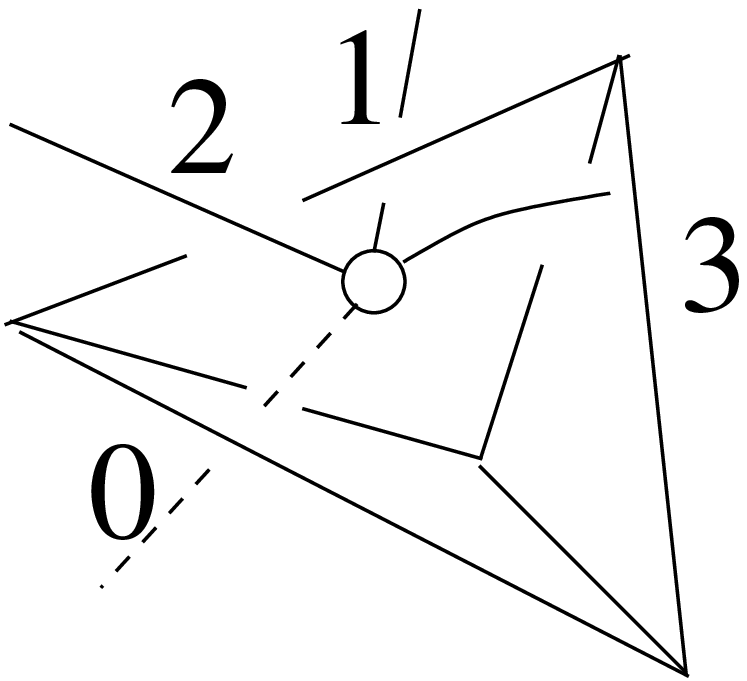,width=0.8cm}\figsubcap{a}}
  \hspace*{15pt}
  \parbox{1.5cm}{\epsfig{figure=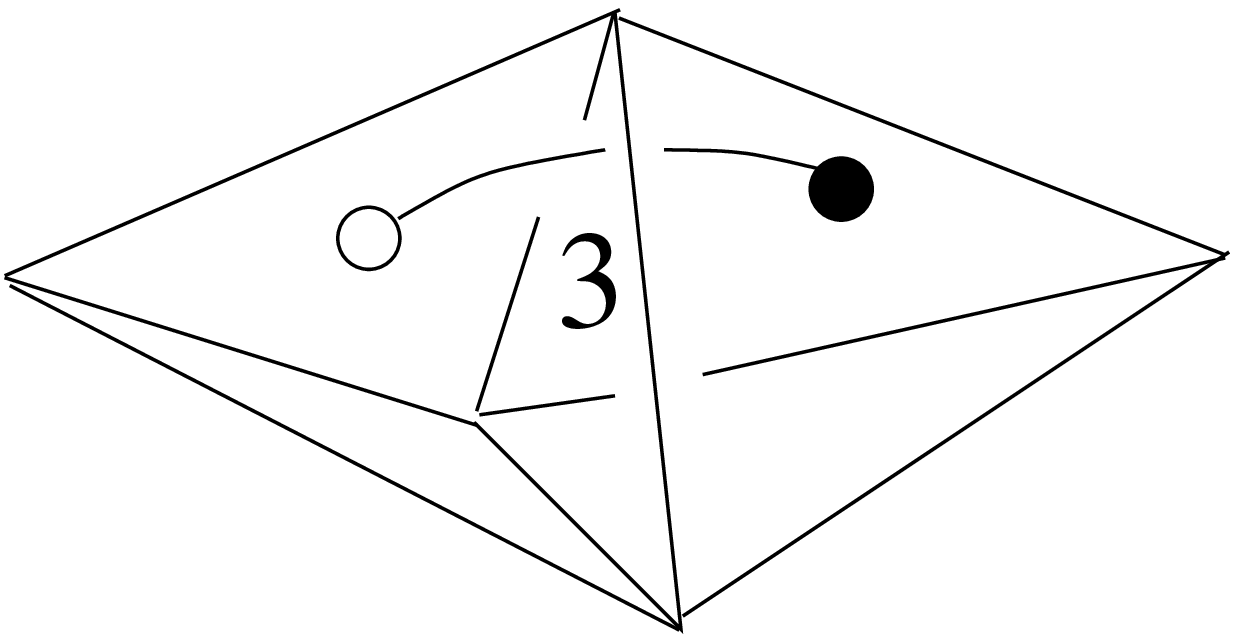,width=1.4cm}\figsubcap{b}}
   \hspace*{15pt}
  \parbox{1.5cm}{\epsfig{figure=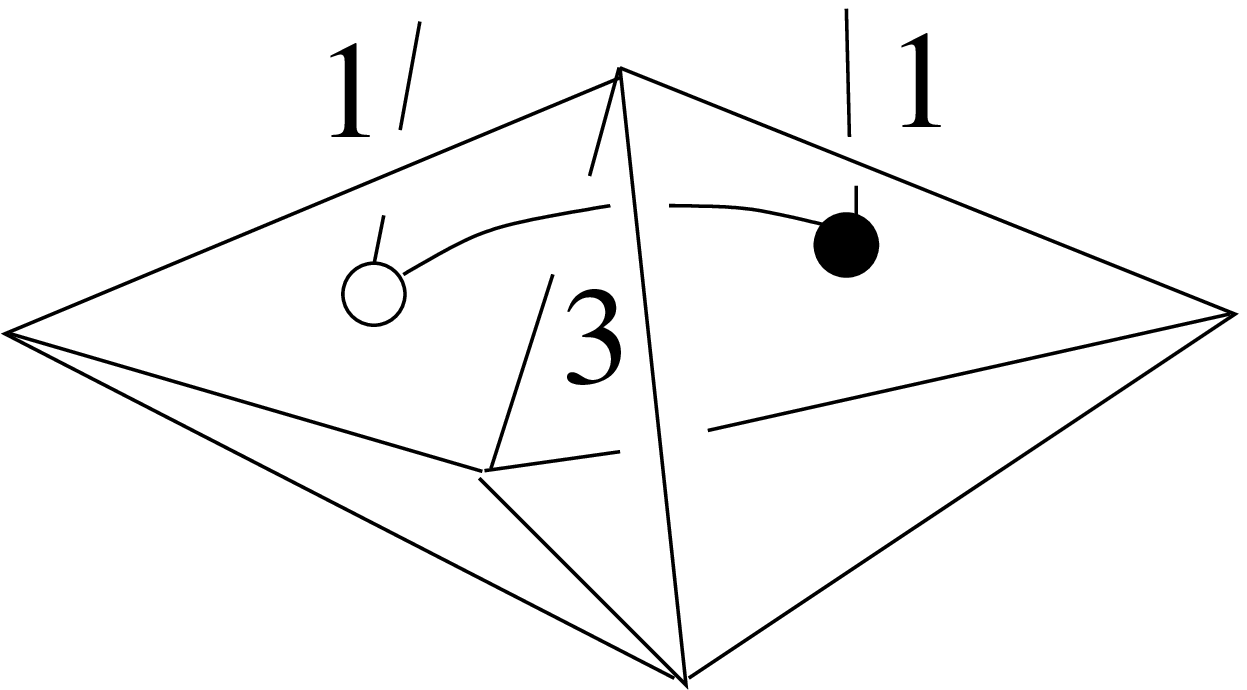,width=1.4cm}\figsubcap{c}}
  \caption{Gluing of tetrahedra associated to a graph. (a) Vertex. (b) Line. (c) Face.}%
  \label{fig1.2}
\end{center}
\end{figure}

Every four valent white (or black) vertex is dual to a positive (or negative) oriented tetrahedron.
The lines emanating from a vertex are dual to the triangles bounding the tetrahedron (see figure \ref{fig1.2}(a)).
The four triangles inherit the color of the lines $0$, $1$, $2$ and $3$. All the lower dimensional simplices 
are then canonically colored: the edge common to the triangles $1$ and $2$ is colored by the couple of colors $12$, and the 
apex common to the triangles $1$, $2$ and $3$ is colored by the triple of colors $123$. 
A line in the graph represents the {\it unique} gluing of two tetrahedra
which respects {\it all} the  colorings. 

Take the example presented in fig. \ref{fig1.2}(b). The line of color $3$ represents the gluing 
of the two tetrahedra along triangles of color $3$ such that the edges $13$, $23$ and $03$ as seen from the positive tetrahedron 
are glued on the edges $13$, $23$ and $03$ as seen from the negative tetrahedron (and similarly for apices).

The cellular structure of the resulting gluing of tetrahedra is encoded in the colors: an edge ($D-2$ simplex for $D+1$ colored graphs)
$13$ corresponds to a subgraph with colors $13$ (see fig. \ref{fig1.2}(c)). Such subgraphs are called the {\it faces} of ${\cal G}$. 

A classical result\cite{Pezzana} guarantees that tensor models generate all manifolds:
\begin{theorem}[Pezzana's Existence Theorem]
 Any piecewise linear $D$ dimensional manifold admits a representation as a $D+1$ colored graph.
\end{theorem}

\begin{figure}[t]
\begin{center}
 \includegraphics[width=2cm]{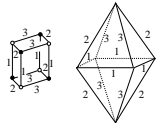}  
\caption{A trace invariant and associated chunk.}
\label{fig:polytope}
\end{center}
\end{figure}

An initial trace invariant (subgraph with colors $1,\dots D$) represents a ``chunk''  of a $D+1$ 
dimensional space. Set $D=3$. As a graph with $3$ colors (see fig. \ref{fig:polytope}) an invariant represents a surface. When seen 
as a subgraph in a $3+1$ colored Feynman graph, the invariant is decorated by lines of color $0$. This amounts to taking the topological 
cone over the surface and obtain the $3+1$ dimensional ``chunk'' of space. Note that, if the surface represented by the invariant 
is non planar, taking the topological cone leads to a conical singularity in the $3+1$ dimensional space.

\paragraph{The $1/N$ expansion.}

The amplitude of a closed connected ribbon graph of a matrix model is\cite{'tHooft:1973jz} $A({\cal G} ) = N^{2-2g({\cal G})}$, where 
$g({\cal G}) $ is the {\it genus} of the graph. What replaces the genus in higher dimensions? We answer this question below.

The genus arises in matrix models because the numbers of vertices, lines and faces in a ribbon graph are not independent. 
A closed connected ribbon graph with  $V=2p$ trivalent vertices has $L = 3p$ lines and
$ F = p + 2 - 2g({\cal G}) $ faces, where $g({\cal G}) $ is the genus of ${\cal G}$.
This generalizes in higher dimensions for $D$ colored graphs. Consider a 
closed connected  $D$ colored graph ${\cal B}$
with $2p$ black and white vertices (and $D p$ lines). The number of faces (subgraphs with two colors)
of ${\cal B}$, is\cite{Gur3,GurRiv,Gur4}
\begin{equation}
   F  = \frac{(D-1)(D-2)}{2} p + (D-1) - \frac{2}{(D-2)!} \omega({\cal B})\; ,
\end{equation}
where the {\it degree} of ${\cal B}$, $\omega({\cal B})$ is a {\it non negative} integer. Of course a similar relation holds 
by shifting $D$ to $D+1$ for $D+1$ colored graphs ${\cal G}$.

\begin{figure}[ht]%
\begin{center}
  \parbox{1.5cm}{\epsfig{figure=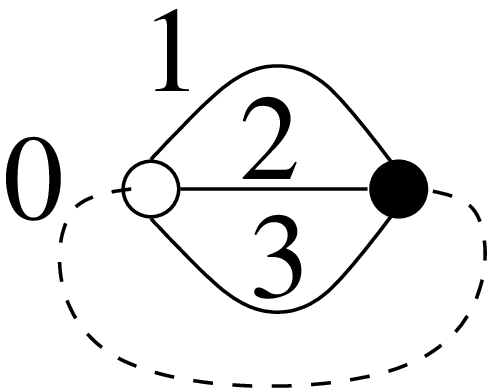,width=1.2cm}\figsubcap{a}}
  \hspace*{8pt}
  \parbox{2.1cm}{\epsfig{figure=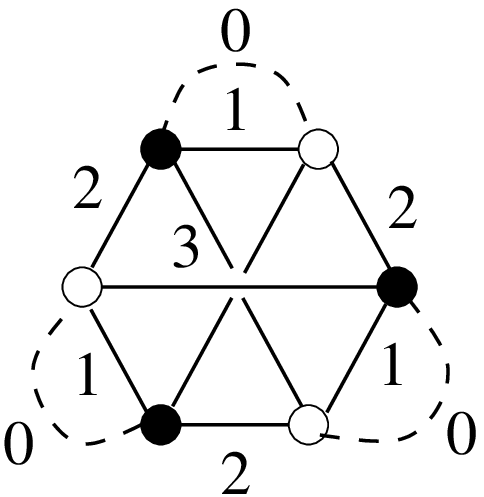,width=1.5cm}\figsubcap{b}}
   \hspace*{8pt}
  \parbox{2.1cm}{\epsfig{figure=tensobsgraph.eps,width=1.8cm}\figsubcap{c}}
  \caption{$3+1$ colored graphs of degree (a) $\omega({\cal G})=0$. (b) $\omega({\cal G})=4$. (c) $\omega({\cal G})=10$.}%
  \label{fig:degree}
\end{center}
\end{figure}

Although, like the genus, the degree is a non negative integer it is {\it not} a topological invariant.
It is an intrinsic number one can compute starting from the graph which mixes information about the 
topology and the cellular structure\cite{Bonzom:2011br} of ${\cal B}$. Some examples of $D+1$ colored graphs and their 
degrees are presented in figure \ref{fig:degree}.

The idea is that, embedded in the graph ${\cal B}$, one can identify some special ribbon graphs ${\cal J}$, called 
jackets\cite{Geloun:2010nw,Gur3,GurRiv,Gur4}.
One then counts the number of faces of each jacket in terms of its genus  $g({\cal J})$. 
Summing over the jackets one obtains the total number of faces of the colored 
graph in terms of the sum of these genera, the degree of the 
colored graph $\omega({\cal B}) = \sum_{{\cal J}} g({\cal J})$.

We chose the scaling of the invariants in the action eq. \eqref{eq:actiongen},
$\omega({\cal B})$, to be precisely the degree of ${\cal B}$. To evaluate 
the amplitude of a Feynman graph ${\cal G}$  (eq. \eqref{eq:ampli}), one needs to count the number of independent sums over indices. 
Each solid line (of colors $1,\dots D$) represents the identification of one index $\delta_{n_i\bar n_i}$. The dashed lines
(of color $0$) represent the identifications of $D$ indices $\prod_{i=1}^D \delta_{a_i \bar p_i}$. An index $n_i$ is 
identified first along a line of color $i$, then along a line of color $0$, then along a line of color $i$
and so on until the cycle of colors $0$ and $i$ closes. We obtain a free sum over an index (hence a factor $N$) for every face 
(subgraph with two colors) of colors $0i$. Combining this with the explicit scalings in eq. \eqref{eq:ampli} we obtain 
the $1/N$ expansion in tensor models in arbitrary dimension:

\begin{theorem}
 The amplitude of a closed connected Feynman ($D+1$ colored) graph generated by the action \eqref{eq:actiongen} is
  \begin{equation}
    A( {\cal G} ) = N^{D - \frac{2}{(D-1)!} \omega( {\cal G} ) }  \; .
  \end{equation}
\end{theorem}

For $D=2$ the $2+1$ colored graphs are ribbon graphs, the degree is the genus and 
the $1/N$ expansion of tensor models reduces to the one of matrices. 

\paragraph{The leading order graphs.}

In matrix models the planar graphs (of genus zero) dominate the $1/N$ expansion. They represent spherical surfaces and form a summable family.
What generalizes planar graphs to arbitrary dimensions? We answer this question below.

The leading order graphs are the $D+1$ colored graphs of degree zero. The structure of 
such graphs is quite different for $D=2$ (matrices) and $D\ge 3$ (tensors). However, for all $D$, the graphs of degree zero 
represent spherical topologies and form a summable family.

\begin{figure}[t]
\begin{center}
 \includegraphics[width=5cm]{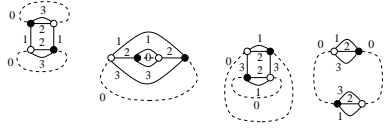}  
\caption{Melons with $p=2$ and $3+1$ colors.}
\label{fig:melonsss}
\end{center}
\end{figure}

A first example of a graph of degree zero is the $D+1$ dipole (see fig. \ref{fig:degree}(a)). 
It has $2$ vertices and $\frac{D(D-1)}{2}$ faces (subgraphs with two colors $ij$) hence degree $0$.
For $D\ge 3$, the $D+1$ colored graphs of degree $0$ with $2p+2$ vertices are obtained by inserting two 
vertices connected by $D$ lines arbitrarily on any line of a $D+1$ colored graph of degree zero with $2p$ vertices\cite{Bonzom:2011zz}. 
This clearly fails for $D=2$.
We call these graphs {\it melons} (see figure \ref{fig:melonsss}). 

The $D+1$ dipole represents a sphere in $D$ dimensions obtained by identifying coherently two $D$ simplices along their boundary.
Two $D+1$ valent vertices connected by $D$-lines represent a $D$ dimensional ball. As the insertion of balls into spheres 
preserves the topology\cite{Gur3,Gur4,GurRiv} we have:
\begin{theorem}
 For any $D$, the graphs of degree $0$ have spherical topology.
\end{theorem}

The melonic graphs are generated by iterative insertions which can be mapped onto abstract (colored, $D+1$-ary) trees.
Melonic graphs thus form a summable family with a finite radius of convergence. They are weighted by the coupling constants
$t_{{\cal B}}$ of the model. When tuning the coupling constants to criticality, tensor models undergo 
a phase transition to infinitely refined continuous random spaces.

\paragraph{The algebra of constraints.}

The observables are graphs with $D$ colors. At leading order only observables corresponding to melonic graphs  ${\cal B}$ contribute.
To each observable we associate a Schwinger Dyson equation\cite{Gurau:2011tj,Gurau:2012ix} and  at leading order only the equations 
corresponding to $D$ melons survive. Each equation translates into a constraint satisfied by the partition 
function ${\cal L}_{\cal T}Z = 0$ for an explicit partial differential operator $ {\cal L}_{\cal T} $. 
The operators form a Lie algebra. As the melons are indexed by $D$ colored trees, the Lie algebra is 
indexed by colored rooted $D$-ary trees\cite{Gurau:2011tj}.

A colored rooted $D$-ary tree ${\cal T }$ is a tree such that all its vertices (including the root) 
have at most $D$ descendants and all its lines have a color index, $1,\dots ,D$. Furthermore the {\it direct  descendants} 
of any vertex are connected by lines with different colors. 

\begin{figure}[ht]%
\begin{center}
  \parbox{2.5cm}{\epsfig{figure=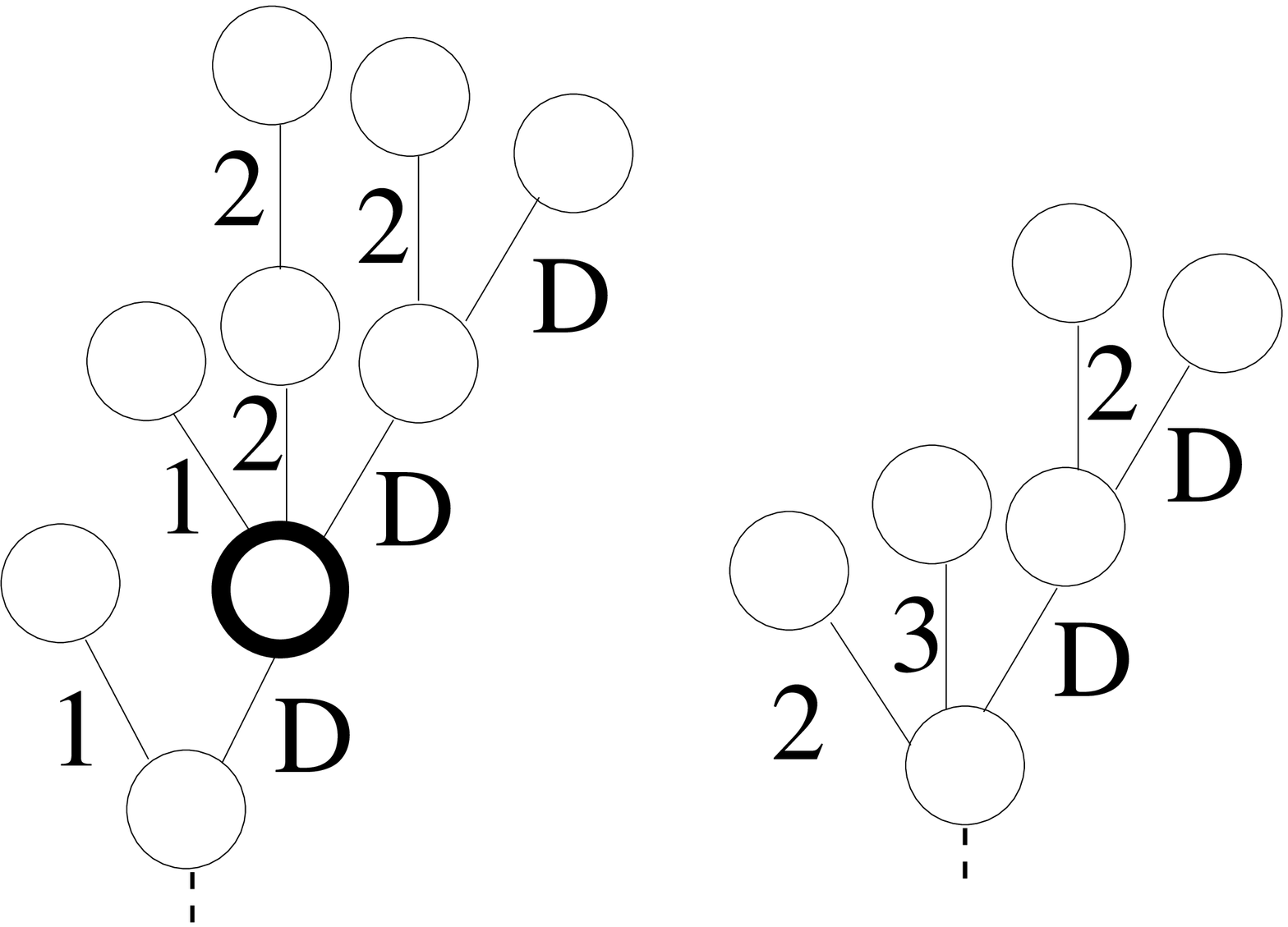,width=2.2cm}\figsubcap{a}}
  \hspace*{15pt}
  \parbox{1.8cm}{\epsfig{figure=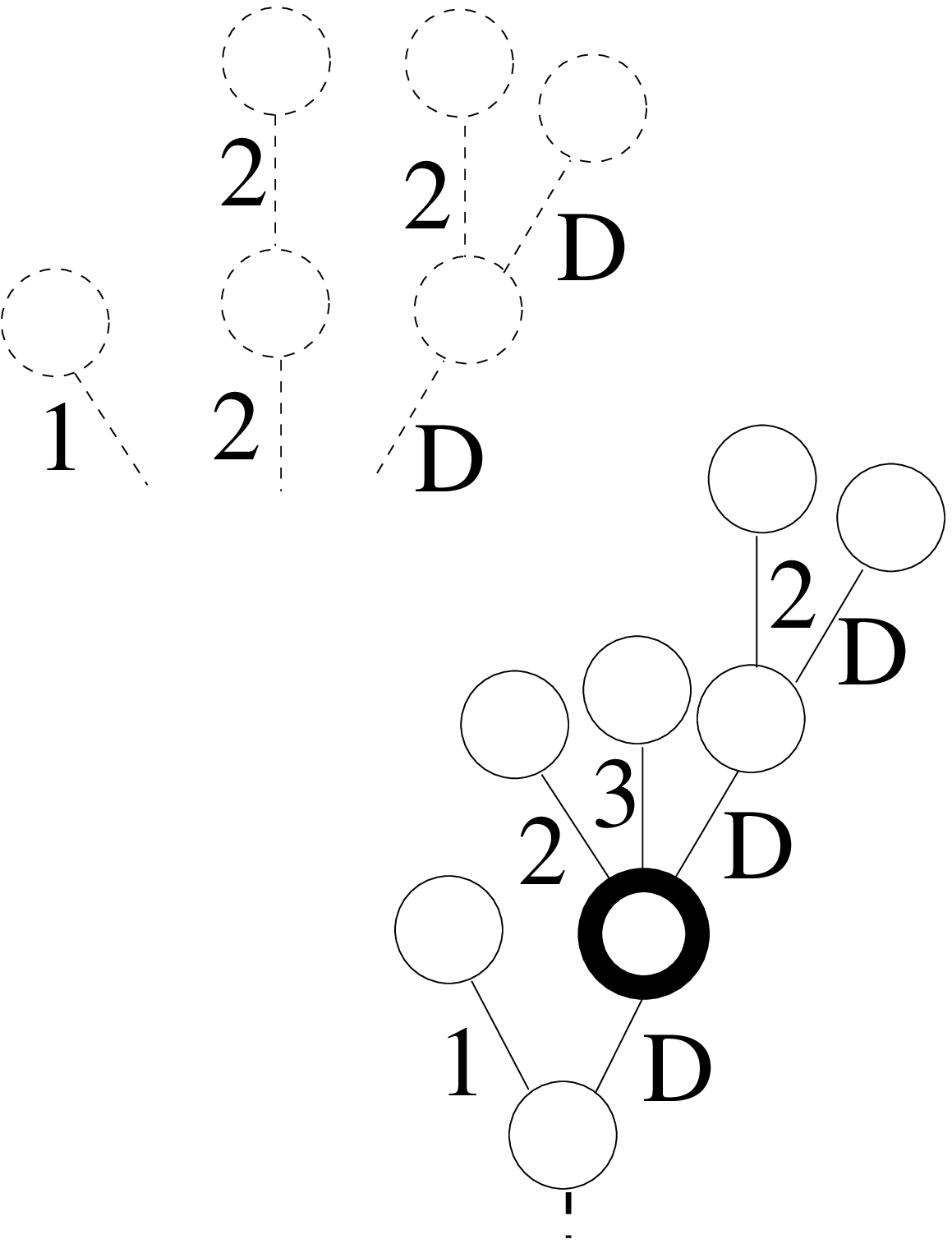,width=1.4cm}\figsubcap{b}}
   \hspace*{15pt}
  \parbox{1.6cm}{\epsfig{figure=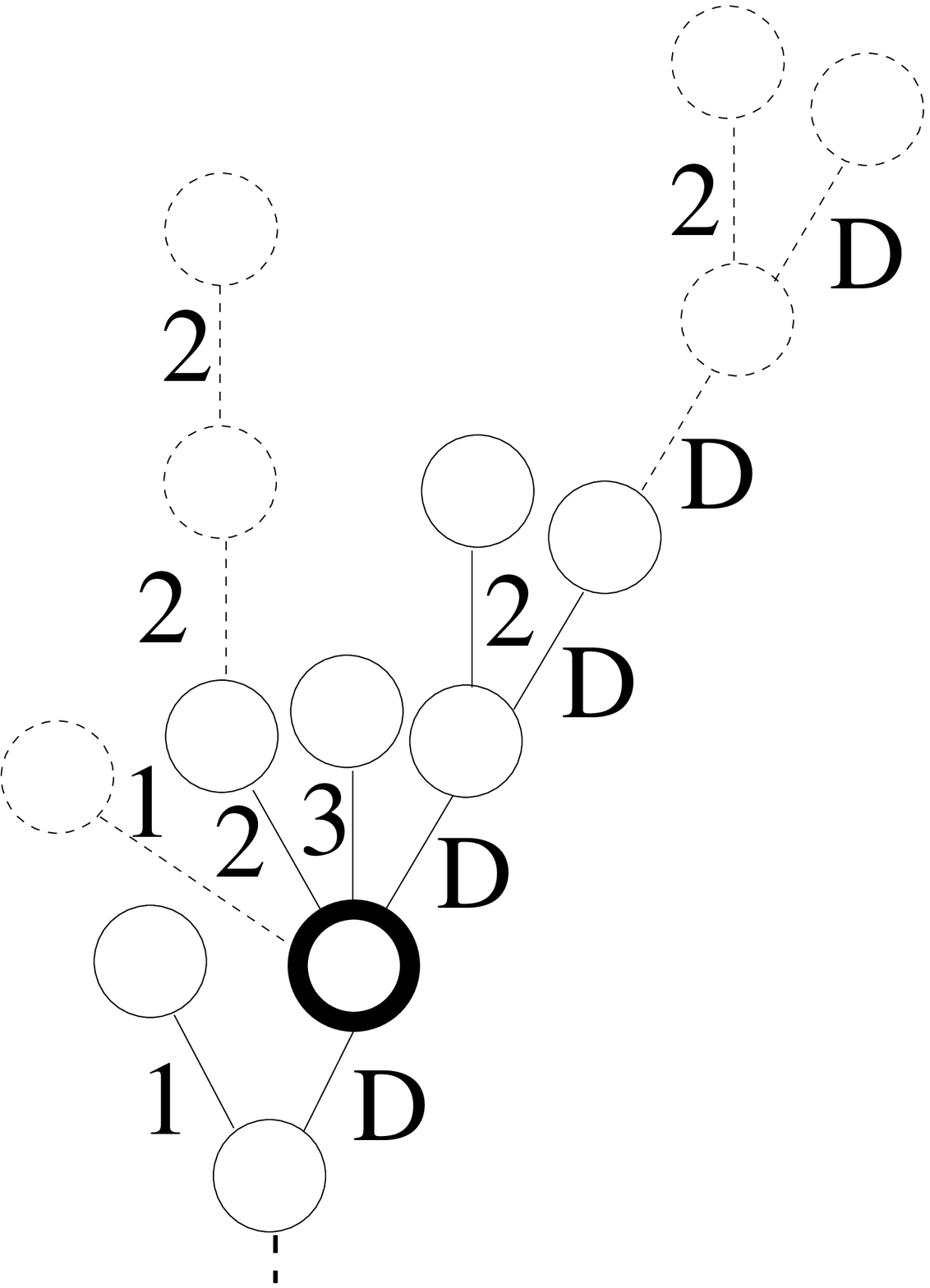,width=1.5cm}\figsubcap{c}}
  \caption{Joining of two trees.}%
  \label{fig:joining}
\end{center}
\end{figure}

Two trees ${\cal T}_1$ and ${\cal T}_2$ can be {\it joined} at a vertex $V\in {\cal T}_1$. The joined tree ${\cal T}_1 \star_V {\cal T}_2$
is obtained as follows (see fig. \ref{fig:joining}, where $V$ is the bold vertex). We first cut the successors of $V$ in $ {\cal T}_1$ (and get 
a collection of branches, represented as dashed in figure \ref{fig:joining}(b)). We then glue the tree ${\cal T}_2$, with its root on 
top of $V$. Finally we reattach the branches at the end of the branches of the appropriate color starting at $V$ (see fig. \ref{fig:joining}(c)).

All the vertices of ${\cal T}_1$ except $V$ and all the vertices in ${\cal T}_2$ except its root, which we denote $(\;)$, are mapped into an unique vertex in
${\cal T}_1 \star_V {\cal T}_2$, while $V$ and $(\; )$ are both mapped into $V$ in ${\cal T}_1 \star_V {\cal T}_2$.
The constraint operators form the Lie algebra with Lie bracket 
\begin{eqnarray}
 \Bigl[ {\cal L}_{{\cal T }_2} , {\cal L}_{{\cal T }_1} \Bigr]   && = \sum_{V \in {\cal T }_2 } {\cal L}_{ {\cal T }_2 \star_V {\cal T }_1 }  -
  \sum_{V \in {\cal T }_1} {\cal L}_{ {\cal T }_1 \star_V {\cal T }_2 }  \; .
\end{eqnarray}

The fact that the bracket is antisymmetric is clear. What is less obvious is that it respects the Jacobi identity. 
All the more so because the composition of trees is {\it not} associative. 
Indeed, $({\cal T}_1 \star_V {\cal T}_2) \star_{W} {\cal T}_3 = {\cal T}_1 \star_V ( {\cal T} \star_{W} {\cal T}_3 ) $
only if $W \in {\cal T}_2$ (or $W =V$). If $W \in {\cal T}_1\setminus V$ the right hand side is not defined.
However the Jacobi identity holds due to the presence of the sums. To see this we evaluate
\begin{eqnarray}
&& \Bigl[ {\cal L}_{{\cal T}_1}, \bigl[ {\cal L}_{ {\cal T}_2 } , {\cal L}_{ {\cal T}_3 } \bigr] \Bigr] =
 \Bigl[ {\cal L}_{{\cal T}_1},  \sum_{V \in {\cal T }_2 } {\cal L}_{ {\cal T }_2 \star_V {\cal T }_3 }  -
  \sum_{V \in {\cal T }_3} {\cal L}_{ {\cal T }_3 \star_V {\cal T }_2 }      \Bigr] \crcr
&& = \sum_{V \in {\cal T }_2 } \Bigl(  \sum_{V' \in {\cal T}_1}   {\cal L}_{{ \cal T}_1 \star_{V'} (  {\cal T }_2 \star_V {\cal T }_3   ) } 
 - \sum_{V' \in { \cal T}_2 \setminus V} {\cal L}_{ (  {\cal T }_2 \star_V {\cal T }_3   ) \star_{V'} {{ \cal T}_1   } }
  \crcr  && \qquad 
    - \sum_{V' \in {\cal T}_3 \setminus (\; ) } {\cal L}_{ (  {\cal T }_2 \star_V {\cal T }_3   ) \star_{V'} {{ \cal T}_1   } }
    -  {\cal L}_{ (  {\cal T }_2 \star_V {\cal T }_3   ) \star_{V} {{ \cal T}_1   } }
\Bigr) \crcr
&& \qquad - \sum_{V \in {\cal T }_3}  \Bigl(  \sum_{V' \in {\cal T}_1}   {\cal L}_{{ \cal T}_1 \star_{V'} (  {\cal T }_3 \star_V {\cal T }_2   ) } 
 - \sum_{V' \in { \cal T}_3 \setminus V} {\cal L}_{ (  {\cal T }_3 \star_V {\cal T }_2   ) \star_{V'} {{ \cal T}_1   } }
  \crcr  && \qquad
    - \sum_{V' \in {\cal T}_2 \setminus (\; ) } {\cal L}_{ (  {\cal T }_3 \star_V {\cal T }_2   ) \star_{V'} {{ \cal T}_1   } }
    -  {\cal L}_{ (  {\cal T }_3 \star_V {\cal T }_2   ) \star_{V} {{ \cal T}_1   } }    \Bigr) \; ,
\end{eqnarray}
hence $\Bigl[ {\cal L}_{{\cal T}_1}, \bigl[ {\cal L}_{ {\cal T}_2 } , {\cal L}_{ {\cal T}_3 } \bigr] \Bigr] +
 \Bigl[ {\cal L}_{{\cal T}_2}, \bigl[ {\cal L}_{ {\cal T}_3 } , {\cal L}_{ {\cal T}_1 } \bigr] \Bigr] +
 \Bigl[ {\cal L}_{{\cal T}_3}, \bigl[ {\cal L}_{ {\cal T}_1 } , {\cal L}_{ {\cal T}_2 } \bigr] \Bigr]   $ writes
\begin{eqnarray}
 &&
  \sum_{V \in {\cal T }_2 }    \sum_{V' \in {\cal T}_1}   {\cal L}_{{ \cal T}_1 \star_{V'} (  {\cal T }_2 \star_V {\cal T }_3   ) } 
 - \sum_{V \in {\cal T }_2 } \sum_{V' \in { \cal T}_2 \setminus V} {\cal L}_{ (  {\cal T }_2 \star_V {\cal T }_3   ) \star_{V'}  {{ \cal T}_1   } }
  \crcr  && \qquad
    - \sum_{V \in {\cal T }_2 } \sum_{V' \in {\cal T}_3 \setminus (\; ) } {\cal L}_{ (  {\cal T }_2 \star_V {\cal T }_3   ) \star_{V'}  {{ \cal T}_1   } }
    - \sum_{V \in {\cal T }_2 } {\cal L}_{ (  {\cal T }_2 \star_V {\cal T }_3   ) \star_{V} {{ \cal T}_1   } }
    \crcr
&& \qquad - \sum_{V \in {\cal T }_3}    \sum_{V' \in {\cal T}_1}   {\cal L}_{{ \cal T}_1 \star_{V'} (  {\cal T }_3 \star_V {\cal T }_2   ) } 
 +\sum_{V \in {\cal T }_3}   \sum_{V' \in { \cal T}_3 \setminus V} {\cal L}_{ (  {\cal T }_3 \star_V {\cal T }_2   ) \star_{V'} {{ \cal T}_1   } }
  \crcr  && \qquad
    +\sum_{V \in {\cal T }_3}  \sum_{V' \in {\cal T}_2 \setminus (\; ) } {\cal L}_{ (  {\cal T }_3 \star_V {\cal T }_2   ) \star_{V'}  {{ \cal T}_1   } }
    +\sum_{V \in {\cal T }_3}   {\cal L}_{ (  {\cal T }_3 \star_V {\cal T }_2   ) \star_{V} {{ \cal T}_1   } }  \crcr
&& +   \sum_{V \in {\cal T }_3 }    \sum_{V' \in {\cal T}_2}   {\cal L}_{{ \cal T}_2 \star_{V'} (  {\cal T }_3 \star_V {\cal T }_1   ) } 
 - \sum_{V \in {\cal T }_3 } \sum_{V' \in { \cal T}_3 \setminus V} {\cal L}_{ (  {\cal T }_3 \star_V {\cal T }_1   ) \star_{V'} {{ \cal T}_2   } }
  \crcr  && \qquad
    - \sum_{V \in {\cal T }_3 } \sum_{V' \in {\cal T}_1 \setminus (\; ) } {\cal L}_{ (  {\cal T }_3 \star_V {\cal T }_1   ) \star_{V'}  {{ \cal T}_2   } }
    - \sum_{V \in {\cal T }_3 } {\cal L}_{ (  {\cal T }_3 \star_V {\cal T }_1   ) \star_{V} {{ \cal T}_2   } }
    \crcr
&& \qquad- \sum_{V \in {\cal T }_1}    \sum_{V' \in {\cal T}_2}   {\cal L}_{{ \cal T}_2 \star_{V'} (  {\cal T }_1 \star_V {\cal T }_3   ) } 
 +\sum_{V \in {\cal T }_1}   \sum_{V' \in { \cal T}_1 \setminus V} {\cal L}_{ (  {\cal T }_1 \star_V {\cal T }_3   ) \star_{V'}  {{ \cal T}_2   } }
  \crcr  && \qquad
    +\sum_{V \in {\cal T }_1}  \sum_{V' \in {\cal T}_3 \setminus (\; ) } {\cal L}_{ (  {\cal T }_1 \star_V {\cal T }_3   ) \star_{V'}  {{ \cal T}_2   } }
    +\sum_{V \in {\cal T }_1}   {\cal L}_{ (  {\cal T }_1 \star_V {\cal T }_3   ) \star_{V} {{ \cal T}_2   } }   \crcr
    && +   \sum_{V \in {\cal T }_1 }    \sum_{V' \in {\cal T}_3}   {\cal L}_{{ \cal T}_3 \star_{V'} (  {\cal T }_1 \star_V {\cal T }_2   ) } 
 - \sum_{V \in {\cal T }_1 } \sum_{V' \in { \cal T}_1 \setminus V} {\cal L}_{ (  {\cal T }_1 \star_V {\cal T }_2   ) \star_{V'} {{ \cal T}_3   } }
  \crcr  && \qquad
    - \sum_{V \in {\cal T }_1 } \sum_{V' \in {\cal T}_2 \setminus (\; ) } {\cal L}_{ (  {\cal T }_1 \star_V {\cal T }_2   ) \star_{V'}{{ \cal T}_3   } }
    - \sum_{V \in {\cal T }_1 } {\cal L}_{ (  {\cal T }_1 \star_V {\cal T }_2   ) \star_{V} {{ \cal T}_3   } }
    \crcr
&& \qquad - \sum_{V \in {\cal T }_2}    \sum_{V' \in {\cal T}_3}   {\cal L}_{{ \cal T}_3 \star_{V'} (  {\cal T }_2 \star_V {\cal T }_1   ) } 
 +\sum_{V \in {\cal T }_2}   \sum_{V' \in { \cal T}_2 \setminus V} {\cal L}_{ (  {\cal T }_2 \star_V {\cal T }_1   ) \star_{V'} {{ \cal T}_3   } }
  \crcr  && \qquad
    +\sum_{V \in {\cal T }_2}  \sum_{V' \in {\cal T}_1 \setminus (\; ) } {\cal L}_{ (  {\cal T }_2 \star_V {\cal T }_1   ) \star_{V'}  {{ \cal T}_3   } }
    +\sum_{V \in {\cal T }_2}   {\cal L}_{ (  {\cal T }_2 \star_V {\cal T }_1   ) \star_{V}  {{ \cal T}_3   } } \; ,
\end{eqnarray}
which cancels due to identities like 
\begin{eqnarray}
&&  - \sum_{V \in {\cal T }_2 } \sum_{V' \in { \cal T}_2 \setminus V} {\cal L}_{ (  {\cal T }_2 \star_V {\cal T }_3   ) \star_{V'}  {\cal L}_{{ \cal T}_1   } }
   + \sum_{V \in {\cal T }_2}   \sum_{V' \in { \cal T}_2 \setminus V} {\cal L}_{ (  {\cal T }_2 \star_V {\cal T }_1   ) \star_{V'} {{ \cal T}_3   } }
   =0 \; , \crcr
&& \sum_{V \in {\cal T }_2 }    \sum_{V' \in {\cal T}_1}   {\cal L}_{{ \cal T}_1 \star_{V'} (  {\cal T }_2 \star_V {\cal T }_3   ) }   
  - \sum_{V \in {\cal T }_1 } \sum_{V' \in {\cal T}_2 \setminus (\; ) } {\cal L}_{ (  {\cal T }_1 \star_V {\cal T }_2   ) \star_{V'} {{ \cal T}_3   } } 
  \crcr
&&     - \sum_{V \in {\cal T }_1 } {\cal L}_{ (  {\cal T }_1 \star_V {\cal T }_2   ) \star_{V}  {{ \cal T}_3   } }
 =    \sum_{V' \in {\cal T}_1}   {\cal L}_{{ \cal T}_1 \star_{V'} (  {\cal T }_2 \star_{(\;)} {\cal T }_3   ) }   
 - \sum_{V \in {\cal T }_1 } {\cal L}_{ (  {\cal T }_1 \star_V {\cal T }_2   ) \star_{V}  {{ \cal T}_3   } } =0 \; , \crcr
 && - \sum_{V \in {\cal T }_3}    \sum_{V' \in {\cal T}_1}   {\cal L}_{{ \cal T}_1 \star_{V'} (  {\cal T }_3 \star_V {\cal T }_2   ) } 
   +\sum_{V \in {\cal T }_1}  \sum_{V' \in {\cal T}_3 \setminus (\; ) } {\cal L}_{ (  {\cal T }_1 \star_V {\cal T }_3   ) \star_{V'}  {{ \cal T}_2   } } 
   \\
 &&   +\sum_{V \in {\cal T }_1}   {\cal L}_{ (  {\cal T }_1 \star_V {\cal T }_3   ) \star_{V} {{ \cal T}_2   } }  =
  -   \sum_{V' \in {\cal T}_1}   {\cal L}_{{ \cal T}_1 \star_{V'} (  {\cal T }_3 \star_{(\;)} {\cal T }_2   ) }
  +\sum_{V \in {\cal T }_1}   {\cal L}_{ (  {\cal T }_1 \star_V {\cal T }_3   ) \star_{V} {{ \cal T}_2   } }  = 0 \; . \nonumber
\end{eqnarray}

This algebra generalizes to all orders to an algebra indexed by $D$-colored graphs\cite{Gurau:2012ix}. 
As it contains the dilation operator\cite{Gurau:2011tj,Gurau:2012ix}, in order to connect with conformal field theories in arbitrary 
dimensions it remains to properly identify the rotation operators. This requires the completion of this Lie algebra by the 
appropriate generalization of the negative part of the Virasoro (Witt) algebra.


\begin{thebibliography}{99}


\bibitem{wishart} J.~Wishart, ``The generalised product moment distribution in samples from a normal multivariate population,''  
    Biometrika {\bf 20 } A, 32, (1928)

\bibitem{wigner} E.~Wigner, ``Characteristic vectors of bordered matrices with infinite dimensions,'' 
   Ann.\ Of\ Math. {\bf 62} (3), 548 (1955). 

\bibitem{DiFrancesco:1993nw}
  P.~Di Francesco, P.~H.~Ginsparg and J.~Zinn-Justin,
  ``2-D Gravity and random matrices,''
  Phys.\ Rept.\  {\bf 254}, 1 (1995)
  [arXiv:hep-th/9306153].

\bibitem{'tHooft:1973jz}
  G.~'t Hooft,
  ``A Planar Diagram Theory for Strong Interactions,''
  Nucl.\ Phys.\ B {\bf 72}, 461 (1974).

\bibitem{Brezin:1977sv}
  E.~Brezin, C.~Itzykson, G.~Parisi and J.~B.~Zuber,
  ``Planar Diagrams,''
  Commun.\ Math.\ Phys.\  {\bf 59}, 35 (1978).

\bibitem{Kazakov:1985ds}
  V.~A.~Kazakov,
  ``Bilocal Regularization of Models of Random Surfaces,''
  Phys.\ Lett.\  B {\bf 150}, 282 (1985).

\bibitem{mm}
  F.~David,
  ``A Model Of Random Surfaces With Nontrivial Critical Behavior,''
  Nucl.\ Phys.\  B {\bf 257}, 543 (1985).

\bibitem{Kazakov:1986hu}
  V.~A.~Kazakov,
  ``Ising model on a dynamical planar random lattice: Exact solution,''
  Phys.\ Lett.\  A {\bf 119}, 140 (1986).

\bibitem{Boulatov:1986sb}
  D.~V.~Boulatov and V.~A.~Kazakov,
  ``The Ising Model on Random Planar Lattice: The Structure of Phase Transition
  and the Exact Critical Exponents,''
  Phys.\ Lett.\  {\bf 186B}, 379 (1987).

\bibitem{Brezin:1989db}
  E.~Brezin, M.~R.~Douglas, V.~Kazakov and S.~H.~Shenker,
  ``The Ising Model Coupled To 2-d Gravity: A Nonperturbative Analysis,''
  Phys.\ Lett.\ B {\bf 237}, 43 (1990).

\bibitem{Kazakovmulticrit}
 V.~A.~Kazakov,
  ``The Appearance of Matter Fields from Quantum Fluctuations of 2D Gravity,''
  Mod.\ Phys.\ Lett.\  A {\bf 4}, 2125 (1989).

\bibitem{Ambjorn:1990ji}
  J.~Ambjorn, J.~Jurkiewicz and Yu.~M.~Makeenko,
  ``Multiloop correlators for two-dimensional quantum gravity,''
  Phys.\ Lett.\  B {\bf 251}, 517 (1990).

\bibitem{Fukuma:1990jw}
  M.~Fukuma, H.~Kawai and R.~Nakayama,
  ``Continuum Schwinger-Dyson Equations and universal structures in 
  two-dimensional quantum gravity,''
  Int.\ J.\ Mod.\ Phys.\  A {\bf 6}, 1385 (1991).

\bibitem{Makeenko:1991ry}
  Yu.~Makeenko,
  ``Loop equations and Virasoro constraints in matrix models,''
  arXiv:hep-th/9112058.

\bibitem{Dijkgraaf:1990rs}
  R.~Dijkgraaf, H.~L.~Verlinde and E.~P.~Verlinde,
  ``Loop equations and Virasoro constraints in nonperturbative 2-D quantum
  gravity,''
  Nucl.\ Phys.\  B {\bf 348}, 435 (1991).


\bibitem{Knizhnik:1988ak}
  V.~G.~Knizhnik, A.~M.~Polyakov and A.~B.~Zamolodchikov,
  ``Fractal Structure of 2D Quantum Gravity,''
  Mod.\ Phys.\ Lett.\  A {\bf 3}, 819 (1988).

\bibitem{david2}
  F.~David,
  ``Conformal Field Theories Coupled to 2D Gravity in the Conformal Gauge,''
  Mod.\ Phys.\ Lett.\ A {\bf 3}, 1651 (1988).

\bibitem{DK}
  J.~Distler and H.~Kawai,
  ``Conformal Field Theory and 2D Quantum Gravity Or Who's Afraid of Joseph Liouville?,''
  Nucl.\ Phys.\ B {\bf 321}, 509 (1989).

\bibitem{Dup}
   Bertrand Duplantier
   ``Conformal Random Geometry"
   Les Houches, Session LXXXIII, 2005, Mathematical Statistical Physics, A. Bovier, F. Dunlop,
   F. den Hollander, A. van Enter and J. Dalibard, eds., pp. 101-217, Elsevier B. V. (2006);
   [arXiv:math-ph/0608053].


\bibitem{double}
  E.~Brezin and V.~A.~Kazakov,
  ``Exactly Solvable Field Theories Of Closed Strings,''
  Phys.\ Lett.\ B {\bf 236}, 144 (1990).

\bibitem{double1}
  M.~R.~Douglas and S.~H.~Shenker,
  ``Strings in Less Than One-Dimension,''
  Nucl.\ Phys.\ B {\bf 335}, 635 (1990).

\bibitem{double2}
  D.~J.~Gross and A.~A.~Migdal,
  ``Nonperturbative Two-Dimensional Quantum Gravity,''
  Phys.\ Rev.\ Lett.\  {\bf 64}, 127 (1990).

\bibitem{ambj3dqg}
  J.~Ambjorn, B.~Durhuus and T.~Jonsson,
  ``Three-Dimensional Simplicial Quantum Gravity And Generalized Matrix
  Models,''
  Mod.\ Phys.\ Lett.\  A {\bf 6}, 1133 (1991).

\bibitem{sasa1}
  N.~Sasakura,
  ``Tensor model for gravity and orientability of manifold,''
  Mod.\ Phys.\ Lett.\  A {\bf 6}, 2613 (1991).

\bibitem{mmgravity}
  M.~Gross,
  ``Tensor models and simplicial quantum gravity in $>$ 2-D,''
  Nucl.\ Phys.\ Proc.\ Suppl.\  {\bf 25A}, 144 (1992).

\bibitem{sasab}
    N.~Sasakura,
  ``Tensor models and hierarchy of n-ary algebras,''
  Int.\ J.\ Mod.\ Phys.\ A {\bf 26}, 3249 (2011),
  arXiv:1104.5312 [hep-th].

\bibitem{sasac}
  N.~Sasakura,
  ``Super tensor models, super fuzzy spaces and super n-ary transformations,''
  Int.\ J.\ Mod.\ Phys.\ A {\bf 26}, 4203 (2011),
  arXiv:1106.0379 [hep-th].

\bibitem{Oriti:2011jm}
  D.~Oriti,
  ``The microscopic dynamics of quantum space as a group field theory,''
  arXiv:1110.5606 [hep-th].

\bibitem{Baratin:2011aa} 
  A.~Baratin and D.~Oriti,
  ``Ten questions on Group Field Theory (and their tentative answers),''
  arXiv:1112.3270 [gr-qc].

\bibitem{color}
  R.~Gurau,
  ``Colored Group Field Theory,''
  Commun.\ Math.\ Phys.\  {\bf 304}, 69 (2011),
  arXiv:0907.2582 [hep-th].

\bibitem{PolyColor}
  R.~Gurau,
  ``Topological Graph Polynomials in Colored Group Field Theory,''
  Annales Henri Poincare {\bf 11}, 565 (2010),
  arXiv:0911.1945 [hep-th].

\bibitem{lost}
  R.~Gurau,
  ``Lost in Translation: Topological Singularities in Group Field Theory,''
  Class.\ Quant.\ Grav.\  {\bf 27}, 235023 (2010),
  arXiv:1006.0714 [hep-th].

\bibitem{coloredreview}
  R.~Gurau and J.~P.~Ryan,
  ``Colored Tensor Models - a review,''
  SIGMA {\bf 8}, 020 (2012)
  [arXiv:1109.4812 [hep-th]].

\bibitem{sefu2}
  J.~B.~Geloun, J.~Magnen and V.~Rivasseau,
  ``Bosonic Colored Group Field Theory,''
  Eur.\ Phys.\ J.\  C {\bf 70}, 1119 (2010),
  arXiv:0911.1719 [hep-th].

\bibitem{Ryan:2011qm}
  J.~P.~Ryan,
  ``Tensor models and embedded Riemann surfaces,''
  Phys.\ Rev.\ D {\bf 85}, 024010 (2012)
  [arXiv:1104.5471 [gr-qc]].

\bibitem{Carrozza:2011jn} 
  S.~Carrozza and D.~Oriti,
  ``Bounding bubbles: the vertex representation of 3d Group Field Theory and the suppression of pseudo-manifolds,''
  Phys.\ Rev.\ D {\bf 85}, 044004 (2012)
  [arXiv:1104.5158 [hep-th]].
 
\bibitem{Carrozza:2012kt} 
  S.~Carrozza and D.~Oriti,
  ``Bubbles and jackets: new scaling bounds in topological group field theories,''
  JHEP {\bf 1206}, 092 (2012)
  [arXiv:1203.5082 [hep-th]].

\bibitem{IsingD}
  V.~Bonzom, R.~Gurau and V.~Rivasseau,
  ``The Ising Model on Random Lattices in Arbitrary Dimensions,''
  arXiv:1108.6269 [hep-th].

\bibitem{EDT}
  D.~Benedetti and R.~Gurau,
  ``Phase Transition in Dually Weighted Colored Tensor Models,''
  Nucl.\ Phys.\ B {\bf 855}, 420 (2012)
  arXiv:1108.5389 [hep-th].

\bibitem{doubletens} 
  R.~Gurau,
  ``The Double Scaling Limit in Arbitrary Dimensions: A Toy Model,''
  arXiv:1110.2460 [hep-th], Phys.\ Rev.\ D {\bf 84}, 124051 (2011)

\bibitem{Bonzom:2012sz} 
  V.~Bonzom,
  ``Multicritical tensor models and hard dimers on spherical random lattices,''
  arXiv:1201.1931 [hep-th].

\bibitem{Bonzom:2012qx} 
  V.~Bonzom and H.~Erbin,
   ``Coupling of hard dimers to dynamical lattices via random tensors,''
  arXiv:1204.3798 [cond-mat.stat-mech].


\bibitem{Gur3}
   R.~Gurau,
  ``The 1/N expansion of colored tensor models,''
  Annales Henri Poincare {\bf 12}, 829 (2011),
  arXiv:1011.2726 [gr-qc].

\bibitem{GurRiv}
  R.~Gurau and V.~Rivasseau,
  ``The 1/N expansion of colored tensor models in arbitrary dimension,''
  Europhys.\ Lett.\  {\bf 95}, 50004 (2011),
  arXiv:1101.4182 [gr-qc].

\bibitem{Gur4}
  R.~Gurau,
  ``The complete 1/N expansion of colored tensor models in arbitrary dimension,''
  Annales Henri Poincare {\bf 13}, 399 (2012)
  [arXiv:1102.5759 [gr-qc]].

\bibitem{Bonzom:2011zz}
  V.~Bonzom, R.~Gurau, A.~Riello and V.~Rivasseau,
  ``Critical behavior of colored tensor models in the large N limit,''
  Nucl. Phys. {\bf B853}, 174-195 (2011),
  arXiv:1105.3122 [hep-th] 

\bibitem{Bonzom:2012hw} 
  V.~Bonzom, R.~Gurau and V.~Rivasseau,
  ``Random tensor models in the large N limit: Uncoloring the colored tensor models,''
  arXiv:1202.3637 [hep-th].

\bibitem{Gurau:2011kk} 
  R.~Gurau,
  ``Universality for Random Tensors,''
  arXiv:1111.0519 [math.PR].

\bibitem{Gurau:2011tj}
  R.~Gurau,
  ``A generalization of the Virasoro algebra to arbitrary dimensions,''
  Nucl.\ Phys.\  B {\bf 852}, 592 (2011)
  [arXiv:1105.6072 [hep-th]].

\bibitem{Gurau:2012ix} 
  R.~Gurau,
  ``The Schwinger Dyson equations and the algebra of constraints of random tensor models at all orders,''
  Nucl.\ Phys.\ B {\bf 865}, 133 (2012)
  [arXiv:1203.4965 [hep-th]].

\bibitem{Bonzom:2012fu} 
  V.~Bonzom,
  ``Revisiting random tensor models at large N via the Schwinger-Dyson equations,''
  arXiv:1208.6216 [hep-th].

\bibitem{BenGeloun:2011rc}
J.~Ben Geloun and V.~Rivasseau,
``A Renormalizable 4-Dimensional Tensor Field Theory,''
arXiv:1111.4997 [hep-th].
  
\bibitem{BenGeloun:2012pu}
  J.~Ben Geloun and D.~O.~Samary,
  ``3D Tensor Field Theory: Renormalization and One-loop $\beta$-functions,''
  arXiv:1201.0176 [hep-th].

\bibitem{BenGeloun:2012yk} 
  J.~Ben Geloun,
  ``Two and four-loop $\beta$-functions of rank 4 renormalizable tensor field theories,''
  arXiv:1205.5513 [hep-th].

\bibitem{Geloun:2012bz} 
  J.~B.~Geloun and E.~R.~Livine,
  ``Some classes of renormalizable tensor models,''
  arXiv:1207.0416 [hep-th].

\bibitem{Carrozza:2012uv} 
  S.~Carrozza, D.~Oriti and V.~Rivasseau,
  ``Renormalization of Tensorial Group Field Theories: Abelian U(1) Models in Four Dimensions,''
  arXiv:1207.6734 [hep-th].

\bibitem{Rivasseau:2011hm}
  V.~Rivasseau,
  ``Quantum Gravity and Renormalization: The Tensor Track,''
  arXiv:1112.5104 [hep-th].

\bibitem{collins}
   B.~Collins, ``Moments and cumulants of polynomial random variables on unitary groups, the Itzykson-Zuber integral, 
   and free probability,'' Int. Math. Res. Not. {\bf 17}, (2003) 953 [arXiv:math-ph/0205010]. 

\bibitem{Pezzana}
   M.~ Pezzana, ``Sulla struttura topologica delle varieta compatte'', Atti\ Sem.\ Mat.\ Fis.\ Univ.\ 
    Modena {\bf 23} (1974), 269.

\bibitem{Bonzom:2011br} 
  V.~Bonzom and M.~Smerlak,
  ``Bubble divergences: sorting out topology from cell structure,''
  Annales Henri Poincare {\bf 13}, 185 (2012)
  [arXiv:1103.3961 [gr-qc]].


\bibitem{Geloun:2010nw} 
  J.~Ben Geloun, T.~Krajewski, J.~Magnen and V.~Rivasseau,
  ``Linearized Group Field Theory and Power Counting Theorems,''
  Class.\ Quant.\ Grav.\  {\bf 27}, 155012 (2010)
  [arXiv:1002.3592 [hep-th]].









\end{thebibliography}
\end{document}